\documentclass[11pt,a4paper]{article}

\usepackage{jheppub}

\usepackage[utf8]{inputenc}
\usepackage{amsmath, amsthm, mathtools}
\usepackage{hyperref}
\usepackage{slashed}

\usepackage{color}
\usepackage{xcolor}
\usepackage{graphicx}
\usepackage{xspace}
\usepackage{wrapfig,enumerate,slashed}

\usepackage[bottom]{footmisc}
\usepackage{wasysym} 
\usepackage{graphicx}
\usepackage{color}
\usepackage{orcidlink}
\usepackage{hyperref}
\usepackage{orcidlink}

\usepackage{subfigure}

\usepackage{soul}

\title{Scalar radiation zeros at the LHC}

\author[a]{Christoph Englert\orcidlink{0000-0003-2201-0667},}
\author[b]{Andrei Lazanu\orcidlink{0000-0002-8061-9828},}
\author[b]{and Peter Millington\orcidlink{0000-0001-6942-8257}}

\affiliation[a]{School of Physics \& Astronomy, University of Glasgow, Glasgow G12 8QQ, United Kingdom}
\affiliation[b]{Department of Physics and Astronomy, University of Manchester, Manchester M13 9PL, United Kingdom}

\emailAdd{christoph.englert@glasgow.ac.uk}
\emailAdd{andrei.lazanu@manchester.ac.uk}
\emailAdd{peter.millington@manchester.ac.uk}

\abstract{
We consider a class of singlet scalar extensions of the Standard Model of particle physics in which the scalar couples only to off-shell states. As a result, low-order tree-level processes involving the singlet scalar vanish, providing a unique phenomenology that may allow to evade existing constraints on new singlet scalar fields. We describe search strategies for such states at the Large Hadron Collider and identify the parameter space that can be explored in the future.
\newline
\newline
This is an author-prepared post-print of \href{https://link.springer.com/article/10.1007/JHEP12(2024)176}{JHEP 12 (2024) 176}, published by Springer Nature under the terms of the \href{https://creativecommons.org/licenses/by/4.0/}{CC BY 4.0} license (funded by SCOAP$^3$).
}


\begin{document}
\maketitle

\section{Introduction}
New physics searches at the Large Hadron Collider (LHC) are well underway but have so far not revealed any concrete signs for physics beyond the Standard Model (BSM). Interpreted from a perspective of large mass gaps between the Standard Model (SM) spectrum and states of its ultraviolet (UV) completion, effective field theory (EFT) methods have seen increasing applications to the phenomenology of hadron collisions, in particular when interpreting results from the LHC. In parallel, the LHC programme needs to safeguard itself from the implicit assumptions that inform EFT approaches. Searches for concrete UV models remain a pillar of the particle-physics programme. Steering away from a priori renormalisable model correlations, signature-driven new physics proposals provide a relevant alternative avenue to pivot and add value to the LHC (and future collider) programme. Most efforts along these lines, so far, have drawn on dark matter, long-lived particles and emerging signature scenarios (see e.g., Ref.~\cite{ATLAS:2021kxv}).

Some of these scenarios are inspired by interactions rooted in cosmological observations~\cite{Brax:2015hma,Burrage:2018dvt,Argyropoulos:2023pmy}, such as the accelerated expansion of the Universe and the unexplained nature of dark energy. 
There, apart from the standard cosmological constant scenario, relying on fine-tuning, significant research has focused on models of scalar fields coupled to gravity. These include Horndeski~\cite{Horndeski1974}, beyond-Horndeski~\cite{Gleyzes:2014dya} and Degenerate Higher Order Scalar-Tensor (DHOST)~\cite{Langlois:2015cwa} theories, which produce second-order equations of motion for the scalar field, avoiding the introduction of ghost instabilities. These scalars are expected to couple to the matter fields of the Standard Model. These couplings can lead to potential signatures in particle colliders that might be detected by current and future colliders~\cite{Brax:2016did}.

In this work, we explore a specific and novel avenue, inspired by these scalar-tensor theories of gravity. Concretely, we identify a subclass of scalar field theory interactions that predominantly manifest themselves through off-shell contributions to physical scattering contributions. Phenomenologically, the associated signatures give rise to a distinct production and decay pattern of the new scalars. Most notably, at leading order, any $1\to 2$ and $2\to 2$ amplitudes involving the scalar and SM matter are identically zero. This is somewhat reminiscent of the well-known SM {\emph{radiation zeros}} in, e.g., gauge-boson pair production (for a review, see Ref.~\cite{Baur:1999ym}), but generalises this phenomenon from the phase space to multi-particle multiplicities by moving away from internal gauge symmetries to specific source terms of the SM Lorentz symmetry currents. The production of such states is then driven by off-shell contributions of Standard Model scattering amplitudes dressed with additional scalar interactions. Through crossing symmetry, the dominant decay proceeds via four-body decay, in stark contrast to any other signature-driven analysis that is currently pursued at the LHC. As production and decay straddle differences in parton luminosity at the LHC, the a priori sensitivity range of displaced-vertex and missing-energy searches is considerably widened. 

This work is organised as follows: In Sec.~\ref{sec:offshell}, we introduce the relevant interactions for a detailed discussion of on-shell zeros in Sec.~\ref{sec:onshell} using an instructive toy example. (We also comment on aspects of higher-order corrections.) We demonstrate how these restrictions are relaxed through off-shell contributions in $2\to 3$ amplitudes, thereby opening up production and detection possibilities at the LHC. The latter are investigated in Sec.~\ref{sec:constraints}, which is devoted to recasting existing and representative searches. We conclude in Sec.~\ref{sec:conc}.

\section{Off-shell matter couplings}
\label{sec:offshell}
To suppress low-order processes involving singlet scalars in extensions of the SM or Einstein gravity, we consider a class of models in which the scalar $\phi$ couples only to the divergence of the SM energy-momentum tensor $T_{\rm SM}^{\mu\nu}$. The matter couplings then take the generic form
\begin{equation}
    \mathcal{L}\supset \left(\partial_{\mu}\partial_{\nu}T_{\rm SM}^{\mu\nu}\right)f(\phi,\partial\phi,\dots)\;,
\end{equation}
where $f(\phi,\partial\phi,\dots)$ is a function of the scalar field and its derivatives. If the Standard Model energy-momentum tensor is conserved on-shell, $\partial_{\mu}T^{\mu\nu}=0$, the scalar $\phi$ can then couple only to off-shell Standard Model degrees of freedom. As we will see in Sec.~\ref{sec:onshell}, this precludes tree-level $t$-channel exchanges of the scalar $\phi$ between Standard Model fermions, avoiding stringent constraints on fifth forces~\cite{Burrage:2017qrf,Ferreira:2019xrr,Brax:2021wcv,Vardanyan:2023jkm,Fischer:2024eic}.

The second divergence of the energy-momentum tensor $\partial_{\mu}\partial_{\nu}T_{\rm SM}^{\mu\nu}$ is a dimension-6 operator, and the lowest-order matter coupling possible is therefore
\begin{equation}
\label{eq:secdiv}
    \mathcal{L}\supset -\frac{C}{M^3}T_{\rm SM}^{\mu\nu}\partial_{\mu}\partial_{\nu}\phi\;,
\end{equation}
where $C$ is a dimensionless constant and $M$ is a mass scale. Equation~\eqref{eq:secdiv} is commonly referred to as longitudinal coupling~\cite{Heisenberg:2014raa}, to be contrasted with the conformal $A(\phi)[T_{\rm SM}]_{\mu}^{\mu}$ and the disformal coupling $B(\phi)\partial_{\mu}\phi\partial_{\nu}\phi T^{\mu\nu}_{\rm SM}$~\cite{Bekenstein:1992pj}.

We note that, in the presence of a spacetime-varying background field $\varphi$, disformal couplings can also generate an operator
\begin{equation}
    \label{eq:vcoupling}
    \mathcal{L}\supset -\frac{C'}{2M^3}T^{\mu\nu}_{\rm{SM}}v_{\{\mu}\partial_{\nu\}}\phi\;,
\end{equation}
where $v_{\mu}=\partial_{\mu}\varphi/M$ and the curly braces indicate symmetrization of the Lorentz indices, i.e., $a^{\{\mu}b^{\nu\}} = a^\mu b^\nu + a^\nu b^\mu$. In Friedmann--Lema\^itre--Robertson--Walker spacetime, one might expect $v_{\mu}=\delta_{\mu}^0v_0$, such that the coupling takes the form
\begin{equation}
    \mathcal{L}\supset -\frac{C'}{2M^3}v_0T^{\{0\mu\}}_{\rm{SM}}\partial_{\mu}\phi\;.
\end{equation}
This Lorentz-violating coupling will be heavily suppressed for a background field that is varying on cosmological timescales. Signatures of Lorentz symmetry violation are investigated at the LHC, see, e.g., the recent analysis targeting a modulation of experimental measurements as a function of sidereal time~\cite{CMS:2024rcv}. These signatures are qualitatively different from ``standard'' LHC searches fundamentally rooted in Lorentz covariance. Equations~\eqref{eq:secdiv} and ~\eqref{eq:vcoupling} also relate to different aspects of the underlying theory. In this work, we will therefore focus on the interactions of Eq.~\eqref{eq:secdiv}, leaving a discussion of Lorentz-violating signatures for future work.

In what follows, we assume that the scalar $\phi$ is canonically normalised, with mass $m_{\phi}$, and vanishing self-interactions. Notice that, in the massless limit $m_{\phi}\to0$, the Lagrangian of the scalar and its matter couplings are shift-symmetric.

\section{Tree-level corrections to massive QED}
\label{sec:onshell}

It is clear from the vanishing of the divergence of the on-shell energy-momentum tensor that the lowest-order tree-level scatterings involving the coupling of $\phi$ to on-shell external states will be zero. Nevertheless, it is illustrative to consider a concrete example. To this end, we focus on the QED Lagrangian for a massive photon
\begin{align}
  \mathcal{L}
  =
  \bar\psi
  \left(
  \frac{i}{2}
  \overleftrightarrow{\slashed{D}}
  -
  m
  \right)
  \psi
  -
  \frac{1}{4} F_{\mu\nu} F^{\mu\nu}
  +
  \frac{1}{2} m_A^2 A_\mu A^\mu
  -
  \frac{1}{2\xi} (\partial_\mu A^\mu)^2
  \;.
\end{align}
Here, $\psi$ is a Dirac fermion of mass $m$, and $A^{\mu}$ is a Proca field of mass $m_A$, with field-strength tensor $F_{\mu\nu}=\partial_{\mu}A_{\nu}-\partial_{\nu}A_{\mu}$. The gauge covariant derivative has been written in the form 
\begin{equation}
\overleftrightarrow{D_{\mu}}=\overleftrightarrow{\partial_{\mu}}-2ie A_{\mu}\;,\qquad f \overleftrightarrow{\partial_\mu} g = f (\partial_\mu g) - (\partial_\mu f) g\;,
\end{equation}
such that the fermion kinetic term is antisymmetrised. The corresponding energy-momentum tensor is~\cite{Freese2022}
\begin{align}
  \label{eqn:emt:qed}
  T^{\mu\nu}
  =
  \frac{i}{4} \bar\psi
  \gamma^{\{\mu} \overleftrightarrow{D}^{\nu\}}
  \psi
  + F^{\mu\sigma} F_{\sigma}^{\phantom{\sigma}\nu}
  + m_A^2 A^\mu A^\nu
  - \frac{1}{\xi} (\partial\cdot A) \partial^{\{\mu} A^{\nu\}}
  -
  \eta^{\mu\nu} \mathcal{L}
  \,.
\end{align}

The couplings~\eqref{eq:secdiv} and~\eqref{eq:vcoupling} lead to the following vertices, wherein all momenta are defined pointing \textit{into} the vertex (e.g., $p_1+p_2+k=0$ in the first expression):
\begin{subequations}
\begin{align}
\label{eq:FeynmanV}
\parbox[c]{3 cm}{\includegraphics[scale=0.7]{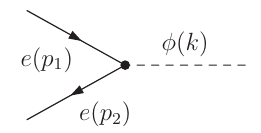}}=&\;i\frac{C}{M^3}\left(p_1+p_2\right)\cdot\left[p_1\left(\slashed{p}_2+m\right)-p_2\left(\slashed{p}_1-m\right)\right],\\
\label{eq:vFeynmanV}
\parbox[c]{3 cm}{\includegraphics[scale=0.7]{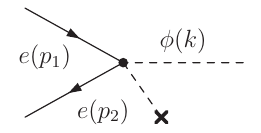}}=&-\frac{C'}{4M^3}\left[\slashed{v}\left(p_1^2-p_2^2\right)\right.\nonumber\\[-0.5cm]&\left.-v\cdot\left(p_1+3p_2\right)\left(\slashed{p}_1-m\right)+v\cdot\left(p_2+3p_1\right)\left(\slashed{p}_2+m\right)\right],\\
\parbox[c]{3 cm}{\includegraphics[scale=0.7]{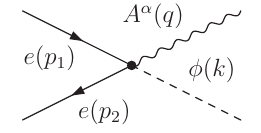}}=& \;i\frac{eC}{M^3}\left(\slashed{k}k_{\mu}-\gamma_{\mu}k^2\right),\label{eq:psipsiAphi}\\
\parbox[c]{3 cm}{\includegraphics[scale=0.7]{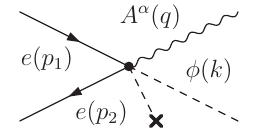}}=&-\frac{eC'}{2M^3}\left(\slashed{v}k_{\mu}+\slashed{k}v_{\mu}-2\gamma_{\mu}v\cdot k\right),\label{eq:psipsiAphiv}\\
\parbox[c]{3.1 cm}{\includegraphics[scale=0.7]{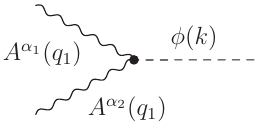}}=&\;\frac{iC}{M^3}\left\{\left[2\left(q_1\cdot k\right)\left(q_2\cdot k\right) -k^2\left(q_1\cdot q_2+m_A^2\right)\right]\eta^{\alpha_1\alpha_2}\right.\nonumber\\[-0.5cm]&+\left(q_1\cdot q_2+m_A^2\right)k^{\alpha_1}k^{\alpha_2}-2\left(q_1\cdot k\right)q_2^{\alpha_1}k^{\alpha_2}-2\left(q_2\cdot k\right)k^{\alpha_1}q_1^{\alpha_2}\nonumber\\&\left.-\xi^{-1}\left[q_1^{\alpha_1}q_2^{\alpha_2}k^2-2\left(q_1\cdot k\right)k^{\alpha_1}q_2^{\alpha_2}-2\left(q_2\cdot k\right)q_1^{\alpha_1}k^{\alpha_2}\right]\right\}\nonumber\\
=&\;\frac{iC}{M^3}\left\{\left[\left(q_1^2+q_2^2-m_A^2\right)\left(q_1\cdot q_2\right)-2q_1^2q_2^2\right]\eta^{\alpha_1\alpha_2}+2\left(q_1\cdot q_2\right)q_1^{\alpha_1}q_2^{\alpha_2}\right.\nonumber\\&-q_1^2q_2^{\alpha_1}\left(q_1+2q_2\right)^{\alpha_2}-q_2^2\left(q_2+2q_1\right)^{\alpha_1}q_1^{\alpha_2}\nonumber\\&+2m_A^2\left(q_1+q_2\right)^{\alpha_1}\left(q_1+q_2\right)^{\alpha_2}+\xi^{-1}\left[q_1^2\left(q_1+2q_2\right)^{\alpha_1}q_2^{\alpha_2}\right.\nonumber\\
&\left.\left.+q_2^2q_1^{\alpha_1}\left(q_2+2q_1\right)^{\alpha_2}+2\left(q_1\cdot q_2\right)\left(q_1^{\alpha_1}q_1^{\alpha_2}+q_2^{\alpha_1}q_2^{\alpha_2}\right)\right]\right\},\label{eq:AAphi}\\
\parbox[c]{3.1 cm}{\includegraphics[scale=0.7]{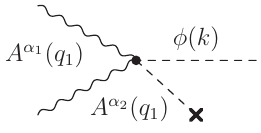}}=&-\frac{C'}{2M^3}\left\{\left[2\left(q_1\cdot v\right)\left(q_2\cdot k\right)+2\left(q_1\cdot k\right)\left(q_2\cdot v\right) \right.\right.\nonumber\\[-0.5cm]&\left.-2\left(v\cdot k\right)\left(q_1\cdot q_2+m_A^2\right)\right]\eta^{\alpha_1\alpha_2}-2\left(q_1\cdot k\right)q_2^{\alpha_1}v^{\alpha_2}-2\left(q_1\cdot v\right)q_2^{\alpha_1}k^{\alpha_2}\nonumber\\&-2\left(q_2\cdot v\right)k^{\alpha_1}q_1^{\alpha_2}-2\left(q_2\cdot k\right)v^{\alpha_1}q_1^{\alpha_2}\nonumber\\&+\left(q_1\cdot q_2+m_A^2\right)\left(v^{\alpha_1}k^{\alpha_2}+k^{\alpha_1}v^{\alpha_2}\right)-2\xi^{-1}\left[q_1^{\alpha_1}q_2^{\alpha_2}\left(v\cdot k\right)\right.\nonumber\\&\left.\left.-\left(q_1\cdot v\right)k^{\alpha_1}q_2^{\alpha_2}-\left(q_1\cdot k\right)v^{\alpha_1}q_2^{\alpha_2}-\left(q_2\cdot v\right)q_1^{\alpha_1}k^{\alpha_2}-\left(q_2\cdot k\right)q_1^{\alpha_1}v^{\alpha_2}\right]\right\}.\label{eq:AAphiv}
\end{align}
\end{subequations}
We have made use of energy-momentum conservation to eliminate the momentum of the $\phi$ field in all but Eqs.~\eqref{eq:psipsiAphi}, \eqref{eq:psipsiAphiv} and \eqref{eq:AAphiv}. Crosses indicate insertion of the constant background vector in the Lorentz-violating vertices. It is readily confirmed that Eqs.~\eqref{eq:vFeynmanV}, \eqref{eq:psipsiAphiv} and \eqref{eq:AAphiv} reduce to Eqs.~\eqref{eq:FeynmanV}, \eqref{eq:psipsiAphi} and \eqref{eq:AAphi}, respectively, in the limit $v_{\mu}\to -ik_{\mu}$, as we would expect.

\begin{figure*}[htbp!]
	\centering    
 \label{fig:ee_to_ee}
 \includegraphics[scale=1]{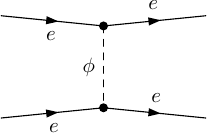}
 \caption{Vanishing $t$-channel exchange of the singlet scalar, which might otherwise give rise to long-range fifth forces for a sufficiently light scalar mediator.
 } 
\end{figure*}

\paragraph{On-shell:} The fermion vertices~\eqref{eq:FeynmanV} and \eqref{eq:vFeynmanV} vanish identically when the fermion four-momenta $p_1$ and $p_2$ are on-shell, i.e., $p_1^2=p_2^2=m^2$, after multiplying from the right by the four-spinor $u(\mathbf{p}_1,s_1)$ and the left by the four-spinor $\bar{v}(\mathbf{p}_2,s_2)$, and making use of the Dirac equations
\begin{subequations}
\begin{align}
\left(\slashed{p}-m\right)u(\mathbf{p},s)=0\;,\\
\bar{v}(\mathbf{p},s)\left(\slashed{p}+m\right)=0\;.
\end{align}
\end{subequations}
This immediately precludes tree-level $t$-channel exchanges of the scalar $\phi$ (see Fig.~\ref{fig:ee_to_ee}) that could give rise to long-range fifth forces.

\begin{figure*}[htbp!]
	\centering    
 \includegraphics[scale=1]{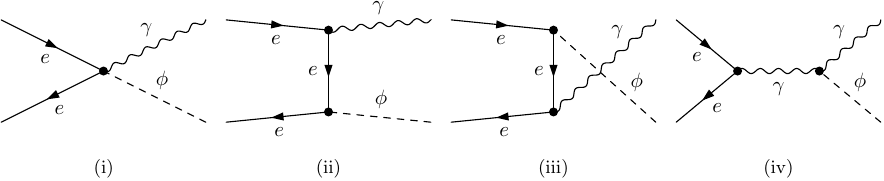}
 \caption{Feynman diagrams for the order-$eC$ process $e^+e^-\to \gamma\phi$, corresponding to the matrix elements in Eq.~\eqref{eq:ee_to_gamma_phi}. These sum to zero, as expected, since the singlet scalar $\phi$ is coupled only to on-shell fermion and photon states. 
 } 
	\label{fig:ee_to_gamma_phi}
\end{figure*}

We notice that the four-point fermion-fermion-vector-scalar vertex does not vanish on-shell. However, this vertex is order $eC$, and we should expect that this cancels against other contributions at order $eC$ on-shell. An example is shown in Figure~\ref{fig:ee_to_gamma_phi}. The four contributions to the matrix element are:
\begin{subequations}
    \label{eq:ee_to_gamma_phi}
    \begin{align}
        i\mathcal{M}_{(i)}&=\overline{v}(\mathbf{p}_2,s_2)\frac{ieC}{M^3}\Big\{\slashed{q}\left(p_1+p_2\right)_{\mu}\nonumber\\&\phantom{=}-2\left[p_1\cdot p_2+q\cdot\left(p_1+p_2\right)+m^2\right]\gamma_{\mu}\Big\}\epsilon^{\mu*}(q)u(\mathbf{p}_1,s_1)\;,\\
        i\mathcal{M}_{(ii)}&=\overline{v}(\mathbf{p}_2,s_2)\frac{ieC}{M^3}\left(p_1\cdot p_2+q\cdot p_2+m^2\right)\gamma_{\mu}\epsilon^{\mu*}(q)u(\mathbf{p}_1,s_1)\;,\\
        i\mathcal{M}_{(iii)}&=\overline{v}(\mathbf{p}_2,s_2)\frac{ieC}{M^3}\left(p_1\cdot p_2+q\cdot p_1+m^2\right)\gamma_{\mu}\epsilon^{\mu*}(q)u(\mathbf{p}_1,s_1)\;,\\
        i\mathcal{M}_{(iv)}&=\overline{v}(\mathbf{p}_2,s_2)\frac{ieC}{M^3}\left[q\cdot\left(p_1+p_2\right)\gamma_{\mu}-\slashed{q}\left(p_1+p_2\right)_{\mu}\right]\epsilon^{\mu*}(q)u(\mathbf{p}_1,s_1)\;,
    \end{align}
\end{subequations}
and we can readily confirm that these sum to zero. This illustrates the delicate cancellations that can occur order by order when there is an underlying symmetry, in this case the conservation of the energy-momentum tensor.

\begin{figure*}[htbp!]
	\centering    
 \includegraphics[scale=1]{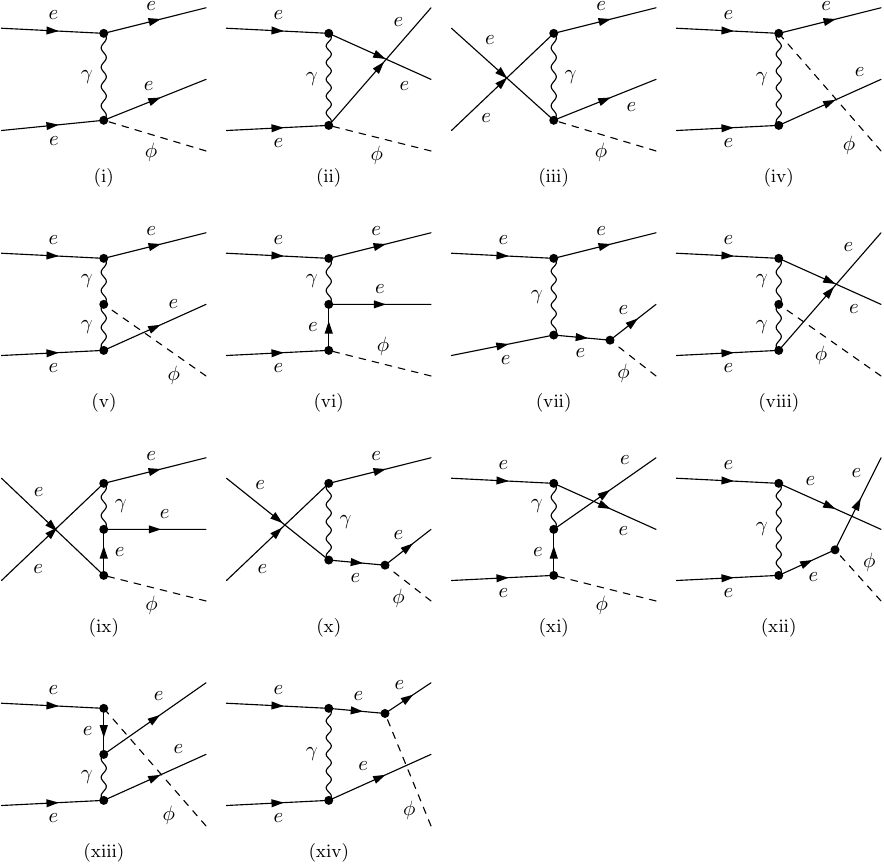}
 \caption{Feynman diagrams contributing to tree-level $e^-e^-\to e^-e^- \phi$ scattering in QED at order~$e^2C$. 
 } 
	\label{fig:ee-eephi}
\end{figure*}

\paragraph{Off-shell:} Continuing with another example from QED, we consider the order $e^2C$ contributions to the $t$-channel and $u$-channel electron-electron scatterings, dressed by a single emission of the scalar, as shown in Fig.~\ref{fig:ee-eephi}. In all amplitudes contributing to this $2\to3$ process, the singlet scalar $\phi$ couples either to at least one off-shell state [Fig.~\ref{fig:ee-eephi} (v)--(xiv)], or through the four-point fermion-fermion-vector-scalar coupling [Fig.~\ref{fig:ee-eephi} (i)--(iv)], which does not vanish (individually) for on-shell states.

Unlike the $2\to 2$ process above, this sum of amplitudes is non-vanishing, and we see that vertices involving the singlet scalar have the interesting property that they only dress processes involving off-shell SM particles, as is expected from the otherwise vanishing divergence of the energy-momentum tensor on-shell. As we see here, this is of particular relevance to higher multiplicity events.

The coupling to off-shell states may also lead to novel couplings of the additional scalar to and via off-shell internal lines in loop-level processes.  Of particular interest are triangle diagrams, which may generate vertices involving SM degrees of freedom that are anomalous with respect to the original symmetries of the model.  This, however, is complicated by the non-renormalizability of this scalar theory, and we leave a comprehensive loop-level analysis to future work.

\section{Constraints from LHC mono-jet analyses}
\label{sec:constraints}
Having reviewed the phenomenological properties of the considered scenario qualitatively above, we now turn to a quantitative discussion within the context of the ongoing LHC physics programme. The absence of direct two- and three-body decays for the scalar results in a phenomenologically distinct and non-common phenomenology of the state. Its decay in four-body decays directly probes the virtuality of the SM matter that is involved in its decay. Compared to other scenarios of long-lived states (such as, e.g., $R$-hadrons), the decay is characterised by a comparably larger phase-space suppression. Therefore, a wider mass range opens up in which the state is stable on collider length scales, as we will discuss below. In this region, it becomes a relevant target of mono-signature analyses. These signatures have been studied by the LHC multi-purpose experiments ATLAS~\cite{ATLAS:2021kxv} and CMS~\cite{CMS:2021far}. Out of all mono-signature channels, mono-jet final states are probed with the highest statistical abundance, and we will focus on these in the following. In these searches, both experiments pursue similar strategies of tagging energetic jets recoiling against ``nothing'' and giving rise to a large missing transverse momentum. As QCD radiation is abundant in hadron colliders at large momentum transfers, there is typically large additional jet activity present in such an event. In order to reduce the contamination from jet energy uncertainty, a typical criterion that is invoked in mono-jet analyses is a significant separation of the recoil system from any other jet in the event.

In the following, we use the basic selection criteria of Ref.~\cite{ATLAS:2021kxv} as a proxy of mono-jet analyses. More specifically, we will employ the inclusive search region of Ref.~\cite{ATLAS:2021kxv}, dubbed `IM0', which is determined by at least one hard jet with
\begin{equation}
p_{T,j} > 150~\text{GeV}
\end{equation}
in $|\eta_j|<4.5$. (Jets are clustered with the anti-kT algorithm with resolution $R=0.4$ and transverse momentum $p_T\geq 30 ~\text{GeV}$.) The IM0 region is characterised by a recoil transverse momentum
\begin{equation}
p_{T,\text{rec}} > 200~\text{GeV}\,.
\end{equation}
Furthermore, the recoil system needs to be sufficiently removed in the azimuthal angle-pseudorapidity plane by $\Delta R > 0.6~(0.4)$ for $ p_{T,\text{rec}} < 250~\text{GeV}$ ($ p_{T,\text{rec}} \geq 250~\text{GeV}$).

\newcommand{\mg}{\texttt{MadGraph5\_aMC@NLO}}

\begin{figure}[!]
\subfigure[\label{fig:excla}]{\includegraphics[height=0.35\textwidth]{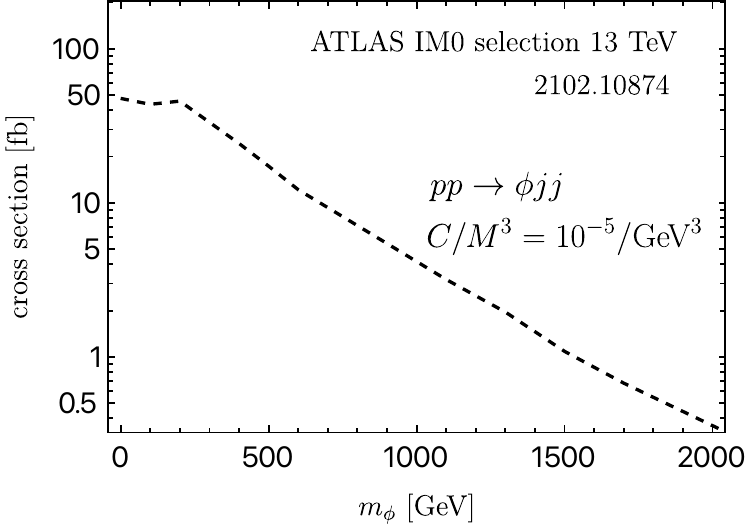}}\hfill
\subfigure[\label{fig:exclb}]{\includegraphics[height=0.35\textwidth]{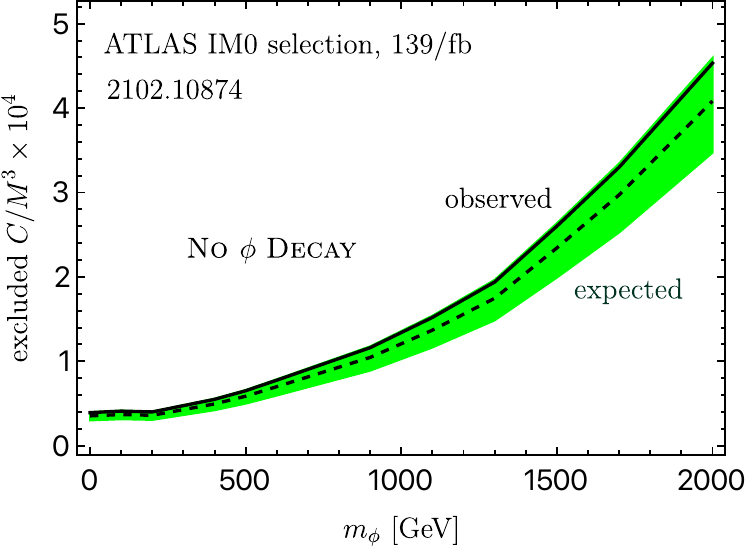}}
\caption{\label{fig:excl} (a) Cross section of the ATLAS IM0 selection for the scalar interaction of Eq.~\eqref{eq:secdiv} before $\phi$ decay for a representative choice of $C/M^3$. These results can be translated into exclusion contours in the $m_\phi-C/M^3$ plane by recasting the results of Ref.~\cite{ATLAS:2021kxv}, as shown in (b) for $\phi$ assumed as stable. The impact $\phi$ lifetime is included in Fig.~\ref{fig:excl2}.}
\end{figure}

We employ \mg~\cite{Alwall:2014hca, Hirschi:2015iia} interfaced with a {\tt{UFO}} model file~\cite{Degrande:2011ua,Darme:2023jdn}, generated with {\tt{FeynRules}}~\cite{Christensen:2008py,Alloul:2013bka} and {\tt{FeynMG}}~\cite{SevillanoMunoz:2022tfb}. Owing to the specific phenomenology associated with $C/M^3$, the first non-trivial hard matrix element to provide a non-zero production cross section is $pp \to \phi j j$,\footnote{Note that other mono-signature searches, such as mono-photon or mono-$Z$ production would necessarily rely on additional jet activity to generate virtuality-driven cross sections. Albeit experimentally cleaner compared to mono-jet searches, these would need to proceed with significant additional jet activity that is typically vetoed in associated searches~\cite{ATLAS:2014wfc,ATLAS:2017nyv,CMS:2017ret} to remove SM backgrounds.} which has been extensively cross checked analytically, numerically, and through the interface to the {\tt{FeynArts}}~\cite{Hahn:2000kx} suite. Events are converted to fully hadronised final states using {\tt{Pythia8}}~\cite{Sjostrand:2007gs,Sjostrand:2014zea,Bierlich:2022pfr}. Reflecting the cut flow of Ref.~\cite{ATLAS:2021kxv} on the fully hadronised final states, we can set limits on the IM0 fiducial region; the cross section for this region and a reference value of $C/M^3$ are shown in Fig.~\ref{fig:excla}. Using the observed, expected and $\pm1\sigma$ regions of Ref.~\cite{ATLAS:2021kxv}, we can translate this cross section into a constraint on $C/M^3$ as a function of the $\phi$ mass when treating this particle as stable (see Fig.~\ref{fig:exclb}).

\begin{figure}[!t]
\subfigure[\label{fig:exclc}]{\includegraphics[height=0.35\textwidth]{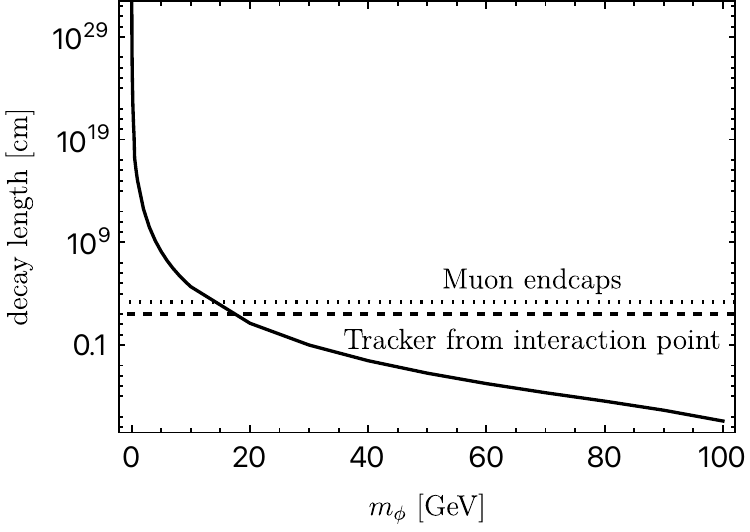}}\hfill
\subfigure[\label{fig:excld}]{\includegraphics[height=0.35\textwidth]{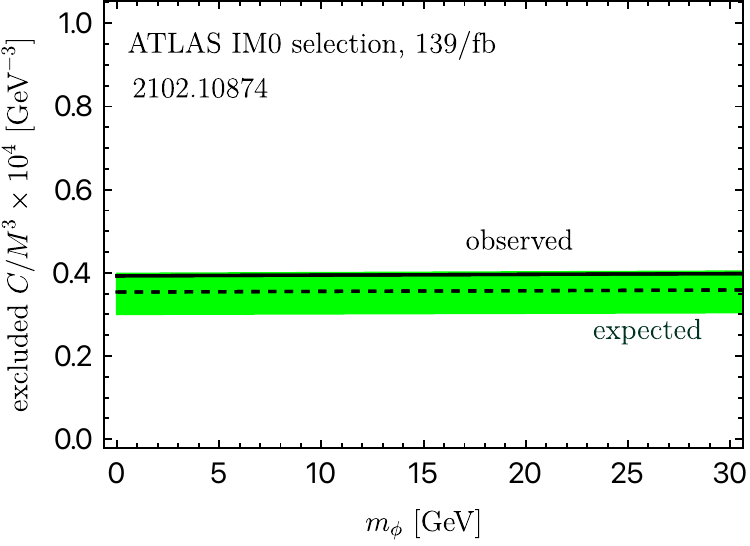}}
\caption{\label{fig:excl2} (a) Decay length of the scalar $\phi$ determined from the ATLAS IM0 selection of Fig.~\ref{fig:exclb}. For a mass range of $m_\phi\lesssim 10~\text{GeV}$, the scalar is stable on collider length scales. (b) The ATLAS exclusion in this region is insensitive to the $m_\phi$ scale.}
\end{figure}

As we have already remarked earlier, the non-zero production of $\phi$ is also linked to a finite lifetime through four-body decays. We can therefore use the results of Fig.~\ref{fig:exclb} to identify the mass region for which the $\phi$ scalar is stable on LHC length scales. The decay width of a particle $\Gamma$ can be linked to the decay length in the LHC lab frame via
\begin{equation}
d= {\beta\over \sqrt{1-\beta^2} }\; {1\over \Gamma}\;.
\end{equation}
If $d$ is comparable to the size of the ATLAS tracker, i.e., $d\leq 1~\text{m}$, the scalar decays before reaching the detector, rendering mono-signature (or displaced-vertex searches) insensitive. On the other hand, if $d$ is large compared to the size of the ATLAS detector itself $d\gtrsim 15~\text{m}$, the particle will escape detection and the mono-jet signature discussed above will be an appropriate selection. Including the exclusion values of $C/M^3$ from Fig.~\ref{fig:exclb}, we can analyse the decay length for the respective values of $m_\phi$ and $C/M^3$ to identify this region. The result is shown in Fig.~\ref{fig:exclc}. For mass scales $m_\phi \lesssim10~\text{GeV}$, the scalar is sufficiently stable to escape detection. In this region, as $m_\phi \ll p_{T,\text{rec}}$, the LHC exclusion determined by the inclusive production of $\phi$ is flat as a function of $m_\phi$ (see Fig.~\ref{fig:excld}). The LHC sensitivity can therefore be estimated as
\begin{equation}
C/M^3 < 4\times 10^{-4}~\text{GeV}^{-3}~\hbox{for}~m_\phi \lesssim 10~\text{GeV}\,.
\end{equation}

In the mass range that indicates a decay length shorter than the tracker, the selection criteria of IM0 (and any missing-energy selection) no longer apply, and the signal stands in competition with a huge (and relatively poorly modelled) QCD multi-jet background. In the intermediate, where the decay length interpolates the different ATLAS detector regions at a reasonable cross section, displaced-vertex~\cite{CMS:2024trg} and emergent signatures~\cite{CMS:2018bvr} become another avenue for detection. However, sensitivity can only be gained in a very narrow parameter window. For instance, sticking to the hard jet criterion of IM0, the requirement of a production cross section of 1~fb with a decay length within the ATLAS experiment equates to a small mass window of $24~\text{GeV}\lesssim m_\phi \lesssim 30~\text{GeV}$. Hence, in the broader context beyond missing energy searches, the model's parameter space also accommodates the displaced vertex signatures that we already mentioned, but also emerging jet signatures~\cite{Schwaller:2015gea}. This signature could be accessible in the mass range of $10 ~\text{GeV} \lesssim m_\phi \lesssim 30~\text{GeV}$ for the selection criteria detailed above. From Fig.~\ref{fig:excld}, it becomes clear that the model in this mass range can accommodate both signatures simultaneously due to the stochastic nature of the particle decay (the decay length is an expectation value). A correlated signature of displaced vertex and emerging jets could, therefore, be a smoking gun of such a scenario, and optimised strategies that lower the jet $p_T$ trigger threshold can enable access to a broader parameter space than suggested by the missing energy signature that we have focused on this work. The precise value of the excluded effective coupling then depends on the optimisation of the analyses as well as experimental efficiencies. These are under continuous review, see e.g. ref. ~\cite{Alimena:2021mdu}; recent proposals employing artificial intelligence for long-lived particle triggering~\cite{Bhattacherjee:2023evs} provide good potential for optimisation. We leave this as an interesting direction for future work.

\section{Conclusions}
\label{sec:conc}

We have analysed a particular derivative coupling of a singlet scalar field to the energy-momentum tensor of the SM degrees of freedom. By virtue of the vanishing of the divergence of the SM energy-momentum tensor on-shell, the singlet scalar field couples only to off-shell states. As a result, standard, tree-level fifth forces are absent, and low-order or low-multiplicity tree-level processes are unaffected. Moreover, this coupling probes the virtuality of the process, leading to a unique phenomenology. This has been illustrated here in the context of mono-jet analyses, where the additional phase-space suppression for the leading four-body decay of the singlet scalar enlarges the parameter space over which the scalar is stable on the scale of the existing multipurpose LHC experiments. By identifying the mass window for which the singlet scalar is stable on these scales, we identify the limits of the sensitivity of mono-signature or displaced vertex searches, which otherwise allow to constrain the strength of the singlet scalar coupling to the SM energy-momentum tensor over this mass window.

We have also identified a class of comparable Lorentz-violating couplings, which also involve only off-shell states. These can arise from disformal couplings to the SM energy-momentum tensor in the case of slowly evolving background fields, as are common in cosmological scenarios. We leave the study of these operators, and the implications of off-shell-only couplings for loop-level processes to future work.

\acknowledgments
All Feynman diagrams presented in this work have been produced using {\tt{FeynArts}}~\cite{Hahn:2000kx} and {\tt{FeynEdit}}~\cite{Hahn:2007ue}. The authors thank Nicolas Chanon, Scott Melville and Sergio Sevillano Mu\~{n}oz for helpful discussions. This work was supported by the Science and Technology Facilities Council (STFC) [Grant No.~ST/X00077X/1] and a United Kingdom Research and Innovation (UKRI) Future Leaders Fellowship [Grant No.~MR/V021974/2]. C.E. is supported by the STFC [Grant No.~ST/X000605/1], the Leverhulme Trust [Research Project Grant RPG-2021-031 and Research Fellowship RF-2024-300\textbackslash9], and the Durham Institute for Particle Physics Phenomenology (IPPP) scheme.

\section*{Data Access Statement}

The analysis presented in this work made use of the following publicly available codes: \texttt{FeynArts} \cite{KUBLBECK1990165}, \texttt{FeynCalc} \cite{Shtabovenko:2020gxv}, \texttt{FormCalc} \cite{Hahn:2016ebn}, \texttt{FeynMG} \cite{SevillanoMunoz:2022tfb}.

\bibliography{references.bib}

\end{document}